\documentclass[preprint,12pt]{elsarticle}
\usepackage{graphicx}
\usepackage{amsthm,amsmath}
\usepackage{amssymb}
\usepackage{mathtools}
\newtheorem{Theorem}{Theorem}[section]

\newtheorem{Remark}{Remark}

\usepackage{mathtools}
\usepackage[toc,page]{appendix}
\usepackage[colorlinks]{hyperref}
\usepackage{bm}
\usepackage{graphicx}
\usepackage{amssymb}
\usepackage{lineno}

\journal{Journal Name}

\usepackage{lineno}
\usepackage{float}
\journal{Journal Name}

\begin{document}
\begin{frontmatter}

\title{A Practical Two-Sample Test for Weighted Random Graphs}

\author[1]{Mingao Yuan\corref{cor1}}
\ead{mingao.yuan@ndsu.edu}

\author[1]{ Qian Wen}
\ead{qian.wen@ndsu.edu }

\cortext[cor1]{Corresponding author}
\address[1]{Department of Statistics, North Dakota State University, Fargo, ND,USA, 58102.}

\begin{abstract}
Network (graph) data analysis is a popular research topic in statistics and machine learning. In application, one is frequently confronted with graph two-sample hypothesis testing where the goal is to test the difference between two graph populations. Several statistical tests have been devised for this purpose in the context of binary graphs. However, many of the practical networks are weighted and existing procedures can't be directly applied to weighted graphs. 
In this paper, we study the weighted graph two-sample hypothesis testing problem and propose a practical test statistic. We prove that the proposed test statistic converges in distribution to the standard normal distribution under the null hypothesis and analyze its power theoretically. The simulation study shows that the proposed test has satisfactory performance and it substantially outperforms the existing counterpart in the binary graph case. A real data application is provided to illustrate the method.
\end{abstract}


\end{frontmatter}

\section{Introduction}
\label{S:1}

A graph or network $\mathcal{G}=(V,E)$ is a mathematical model that consists of a set $V$ of nodes (vertices)  and a set $E$ of edges. In the last decades, it has been widely used to represent a variety of systems in various regimes \cite{N04,COTR11,F10, M11, CY06}. For instance, in social networks, a node denotes an individual and an edge represents the interaction between two individuals \cite{F10}; in brain graphs, a node may be a neural unit and the functional link between two units forms an edge \cite{GAMVB18}; in co-authorship networks, the authors of a collection of articles are the nodes and an edge is defined to be the co-authorship of two authors \cite{N04}. Due to the widespread applications, network data analysis has drawn a lot of attentions in both statistical and machine learning communities \cite{A18, ABB06,ACB13,BS16,GL17,L16,YN20}. Most of the existing literature focus on mining a single network, such as community detection \cite{A18,ABB06,BS16,ACB13}, global testing of the community structures \cite{BS16,L16,GL17,YN20} and so on. In practice, a number of graphs from multiple populations may be available. For example, in the 1000 Functional Connectomes Project, 1093 fMRI (weighted) networks were collected from subjects located in 24 communities \cite{CLBRK17}; to study the relation between Alzheimer's disease and a functional disconnection of distant brain areas, dozens of functional connectivity (weighted) networks from patients and control subjects were constructed \cite{SJNBS07}. In this case, a natural and fundamental question is to test the difference between two graph populations, known as graph two-sample hypothesis testing. 

There are a few literature dealing with the graph two-sample hypothesis testing problem \cite{CLBRK17,TASLP17,DMAU19,GL18}. Specifically, \cite{CLBRK17} firstly investigated this problem and proposed a $\chi^2$-type test. In \cite{TASLP17}, the authors developed a kernel-based test statistic for random dot product graph models. Under a more general setting, \cite{DMAU19} studied the graph two sample test from a minimax testing perspective and proposed testing procedures based on graph distance such as Frobenius norm or operator norm. The threshold of the test statistics in \cite{DMAU19} could be calculated by concentration inequalities, which usually makes the test very conservative \cite{GL18}. To overcome this issue, \cite{GL18} derived the asymptotic distribution of the test statistic and proposed practical test methods that outperform existing methods.

In practice, most of the graphs are weighted \cite{CLBRK17,SJNBS07,TB11,AJC15,A14}. The testing procedures in \cite{GL18,DMAU19,TASLP17, CZL20} are designed under the context of binary (unweighted) graphs and the tests can't be directly applied to weighted graphs (See Section \ref{simu} for an example). Consequently, before using these tests, one has to artificially convert weighted graphs into binary graphs, which can result in a loss of information \cite{TB11,AJC15,A14}. Motivated by the $T_{fro}$ test in \cite{GL18}, we propose a powerful test statistic for weighted graph two-sample hypothesis testing. Under the null hypothesis, the proposed test statistic converges in distribution to the standard normal distribution and the power of the test is theoretically characterized. Simulation study shows that the test can achieve high power and it substantially outperforms its counterpart in the binary graph case. Besides, we apply the proposed test to a real data.

The rest of the paper is organized as follows. In Section \ref{twosampletest}, we formally state the weighted graph two-sample hypothesis testing problem and present the theoretical results. In Section \ref{simu}, we present the simulation study results and real data application. The proof of main result is deferred to Section \ref{proof}.

\section{Weighted Graph Two-Sample Hypothesis Test}\label{twosampletest}
For convenience, let $X\sim F$ represent random variable $X$ follows distribution $F$ and let $Bern(r)$ denote the Bernoulli distribution with success probability $r$. 

Let $V=\{1,2,\dots,n\}$ be a vertex (node) set and $\mathcal{G}=(V,E)$ denote an undirected graph on $V$ with edge set $E$. The adjacency matrix of graph $\mathcal{G}$ is a symmetric matrix $A\in\{0,1\}^{n\times n}$ such that $A_{ij}=1$ if $(i,j)\in E$ and 0 otherwise. The graph $\mathcal{G}$ is binary or unweighted, since $A_{ij}$ only records the existence of an edge. If $A_{ij}\sim Bern(p_{ij}), 0\leq p_{ij}\leq1$, then the graph  $\mathcal{G}$ is called an inhomogeneous random graph(inhomogeneous Erd\"{o}s-R\'{e}nyi graph). Let $\mu=(\mu_{ij})_{1\leq i<j\leq n}$ be a sequence of real numbers and $Q=(Q_{ij})_{1\leq i<j\leq n}$ be a sequence of distributions defined on a bounded interval, where each $Q_{ij}$ is uniquely parametrized by its mean value $\mu_{ij}$.
A weighted random graph $\mathcal{G}=(V,Q,\mu)$ is defined as follows.
 For nodes $i,j$, 
\[A_{ij}=A_{ji},\ \ A_{ij}\sim Q_{ij}(\mu_{ij}),\ 1\leq i<j\leq n,\]
$A_{ii}=0$ $(i=1,2,\dots,n)$ and $A_{ij}$ is independent of $A_{kl}$ if $\{i,j\}\neq \{k,l\}$. If $Q_{ij}(\mu_{ij})=Bern(\mu_{ij})$, then $\mathcal{G}=(V,Q,\mu)$ is just the inhomogeneous random graph \cite{GL18,DMAU19}.

Given i.i.d. graph sample $G_1,\dots,G_m\sim \mathcal{G}_1=(V,Q,\mu_1)$ and i.i.d. graph sample $H_1,\dots,H_m\sim \mathcal{G}_2=(V,Q,\mu_2)$, we are interested in the weighted graph two-sample hypothesis testing problem
\begin{equation}\label{hypo}
H_0: \mathcal{G}_1=\mathcal{G}_2,\hskip 1cm H_1: \mathcal{G}_1\neq\mathcal{G}_2.
\end{equation}
Let $A_{G_k}$ and $A_{H_k}$ be the adjacency matrix of graph $G_k$ and $H_k$ respectively. Then $A_{G_k,ij}\sim Q_{ij}(\mu_{1,ij})$ and $A_{H_k,ij}\sim Q_{ij}(\mu_{2,ij})$ for $1\leq i<j\leq n$. Consequently, (\ref{hypo}) is equivalent to the following hypothesis test
\[H_0: \mu_1=\mu_2,\hskip 1cm H_1: \mu_1\neq\mu_2.\]

In the binary graph case ($Q_{ij}(\mu_{ij})=Bern(\mu_{ij}),1\leq i<j\leq n$), several testing procedures for (\ref{hypo}) are available in the literature. 
For $m\rightarrow\infty$ and small $n$, a $\chi^2$-type test was proposed in \cite{CLBRK17}. For $m=1$ and $n\rightarrow\infty$, under the random dot product model, a nonparametric test statistic was developed in \cite{TASLP17}, and a test based on eigenvalues of adjacency matrix under the inhomogeneous random graph could be found in \cite{GL18}. A more practical case is small $m$ $(m\geq2)$ and $n\rightarrow\infty$. In this case, a test called $T_{fro}$ was proposed in \cite{DMAU19} and its asymptotic behavior was studied in \cite{GL18}. Recently, \cite{CZL20} proposed a test statistic based on the largest eigenvalue of a Wigner matrix.

In this work, we study (\ref{hypo}) for a broad class of distributions $Q$ and focus on the regime $m\geq2$ and $n\rightarrow\infty$. The sample size $m$ could be either fixed or tend to infinity along with $n$.

 To define the test statistic, the two samples $G_k,H_k, (1\leq k\leq m)$ are randomly partitioned into two parts, denoted as $G_k,H_k, (1\leq k\leq m/2)$ and $G_k,H_k, (m/2< k\leq m)$ with a little notation abuse. Let $s_n^2=\sum_{1\leq i<j\leq n}T_{ij}^2$, where
\[T_{ij}=\sum_{k\leq \frac{m}{2}}(A_{G_k,ij}-A_{H_k,ij})\sum_{k> \frac{m}{2}}(A_{G_k,ij}-A_{H_k,ij}).\]
We propose the following test statistic for (\ref{hypo}):
\begin{equation}\label{tstat}
\mathcal{T}_n=\frac{\sum_{1\leq i<j\leq n}T_{ij}}{s_n}.
\end{equation}

Let $\sigma_{ij}^2=Var(A_{G_k,ij})$ and $\eta_{ij}=\mathbb{E}(A_{G_k,ij}-A_{H_k,ij})^4$ under $H_0$. The asymptotic distribution of $\mathcal{T}_n$ is given in the following theorem.

\begin{Theorem}\label{thmel}
Suppose 
\begin{eqnarray}\nonumber
&&n=o\big(\sum_{1\leq i<j\leq n}\sigma_{ij}^4\big),\ \hskip 1.2cm \ \frac{\sum_{1\leq i<j\leq n}\sigma_{ij}^8}{(\sum_{1\leq i<j\leq n}\sigma_{ij}^4)^2}=o(1), \ \\ \label{cond} 
&&\frac{\sum_{1\leq i<j\leq n}\sigma_{ij}^4\eta_{ij}}{m(\sum_{1\leq i<j\leq n}\sigma_{ij}^4)^2}=o(1),\ \ \frac{\sum_{1\leq i<j\leq n}\eta_{ij}^2}{m^2(\sum_{1\leq i<j\leq n}\sigma_{ij}^4)^2}=o(1).
\end{eqnarray}
Then under $H_0$, 
$\mathcal{T}_n$ converges in distribution to $N(0,1)$, the standard normal distribution, as $n\rightarrow\infty$.
\end{Theorem}

Theorem \ref{thmel} states that the limiting distribution of $\mathcal{T}_n$ under the null hypothesis is $N(0,1)$. Given type one error $\alpha$,
reject $H_0$ if $|\mathcal{T}_n|>Z_{(1-\frac{\alpha}{2})}$ where $Z_{(1-\frac{\alpha}{2})}$ is the $100(1-\frac{\alpha}{2})\%$ quantile of the standard normal distribution.

Condition (\ref{cond}) could be simplified in the binary case. 
Suppose $Q_{ij}(\mu_{ij})=Bern(\mu_{ij})$ and $\mu_{1,ij}=\mu_{2,ij}=\mu_{ij}\leq 1-\delta$ for some $
\delta\in(0,1)$ under $H_0$. Then $\eta_{ij}\leq2\sigma_{ij}^2\leq2\mu_{ij}$. In this case, condition (\ref{cond}) reduces to $n=o(\|\mu\|_F^2)$. Here $\|\mu\|_F$ denotes the Frobenius norm of matrix $\mu$. To see this, let $C$ be a generic constant, then \begin{equation}\label{cond2}
0.5\delta^2\|\mu\|_F^2=\delta^2 \sum_{1\leq i<j\leq n}\mu_{ij}^2\leq \sum_{1\leq i<j\leq n}\sigma_{ij}^4\leq \sum_{1\leq i<j\leq n}\mu_{ij}^2=0.5\|\mu\|_F^2,
\end{equation}
\begin{equation}\label{cond3}
\frac{\sum_{1\leq i<j\leq n}\sigma_{ij}^8}{(\sum_{1\leq i<j\leq n}\sigma_{ij}^4)^2}\leq C\frac{\sum_{1\leq i<j\leq n}\mu_{ij}^4}{(\sum_{1\leq i<j\leq n}\mu_{ij}^2)^2}\leq C\frac{\sum_{1\leq i<j\leq n}\mu_{ij}^2}{(\sum_{1\leq i<j\leq n}\mu_{ij}^2)^2}=C\frac{1}{\|\mu\|_F^2}\rightarrow0, 
\end{equation}
\begin{equation}\label{cond4}
\frac{\sum_{1\leq i<j\leq n}\sigma_{ij}^4\eta_{ij}}{m(\sum_{1\leq i<j\leq n}\sigma_{ij}^4)^2}\leq C \frac{\sum_{1\leq i<j\leq n}\mu_{ij}^2}{m(\sum_{1\leq i<j\leq n}\mu_{ij}^2)^2}=C\frac{1}{m\|\mu\|_F^2}\rightarrow0,
\end{equation}
\begin{equation}\label{cond5}
\frac{\sum_{1\leq i<j\leq n}\eta_{ij}^2}{m^2(\sum_{1\leq i<j\leq n}\sigma_{ij}^4)^2}\leq C\frac{\sum_{1\leq i<j\leq  n}\mu_{ij}^2}{m^2(\sum_{1\leq i<j\leq n}\mu_{ij}^2)^2} =C\frac{1}{m^2\|\mu\|_F^2}\rightarrow0.
\end{equation}
If $n=o(\|\mu\|_F^2)$, then (\ref{cond}) holds by (\ref{cond2}),(\ref{cond3}),(\ref{cond4}),(\ref{cond5}).

In the following, we analyze the power of the proposed test statistic.
Let $\sigma_{1,ij}^2=Var(A_{G_k,ij})$, $\sigma_{2,ij}^2=Var(A_{H_k,ij})$ under $H_1$ and
\[ V_{ij}=\sigma_{1,ij}^2+\sigma_{2,ij}^2+(\mu_{1,ij}-\mu_{2,ij})^2,\ \ \lambda_n=\frac{m\sum_{1\leq i<j\leq n}(\mu_{1,ij}-\mu_{2,ij})^2}{2\sqrt{\sum_{1\leq i<j\leq n}V_{ij}^2}}.\]
\begin{Theorem}\label{thme2}
Suppose $n=o\big(m\sum_{1\leq i<j\leq n}V_{ij}^2\big)$. Then under $H_1$, 
$\mathcal{T}_n=\lambda_n+O_P(1)$.
\end{Theorem}
According to Theorem \ref{thme2},
the power of the test goes to one if
$\lambda_n\rightarrow\infty$, as $n\rightarrow\infty$.
 The expression of $\lambda_n$ explicitly characterizes the effect of sample size $m$ and the mean and variance of edge weight on the power of the test statistic.  

In the following, let's restrict Theorem \ref{thme2} to binary graphs to see when the test could achieve high power.
Suppose $Q_{ij}(\mu_{t,ij})=Bern(\mu_{t,ij})$, $\mu_{t,ij}\rightarrow0$, and $\mu_{1,ij}/ \mu_{2,ij}\rightarrow\tau$($\tau>0$) for $t=1,2$. Then
\[V_{ij}=\mu_{1,ij}(1-\mu_{1,ij})+\mu_{2,ij}(1-\mu_{2,ij})+(\mu_{1,ij}-\mu_{2,ij})^2=(\mu_{1,ij}+\mu_{2,ij})(1+o(1)).\]
In this case, $n=o(m\sum_{1\leq i<j\leq n}V_{ij}^2)$ requires $n=o(m\|\mu_1+\mu_2\|_F^2)$. Besides,
\begin{equation}\label{lambda}
\lambda_n=\frac{m\sum_{1\leq i<j\leq n}(\mu_{1,ij}-\mu_{2,ij})^2}{2\sqrt{\sum_{1\leq i<j\leq n}(\mu_{1,ij}+\mu_{2,ij})^2}}(1+o(1))=(1+o(1))\frac{m\|\mu_1-\mu_2\|_F^2}{2\sqrt{2}\|\mu_1+\mu_2\|_F}.
\end{equation}
For fixed $\mu_1$ and $\mu_2$, as the sample size $m$ increases, the power increases. 
As $\|\mu_1-\mu_2\|_F^2$ gets larger, the power gets higher when $\|\mu_1+\mu_2\|_F^2$ and $m$ are held constant.

\begin{Remark}\label{remark1}
The quantity $\lambda_n$ in Theorem \ref{thme2} completely characterizes the power of our test. For binary graphs, the sparsity may increase or decrease the power, dependent on the model settup. To see this, we consider two scenarios below.\\

(a) Suppose $\mu_{1,ij}=\tau a_n$ for a constant $\tau>0$ and $\mu_{2,ij}=a_n$ with $a_n=o(1)$, $1\leq i<j\leq n$. By (\ref{lambda}), it follows that
\[\lambda_n=mna_n\frac{(\tau-1)^2}{4(\tau+1)}[1+o(1)].\]
For fixed sample size $m$ and the number of nodes $n$, the power of our test statistic declines as the networks get sparser (smaller $a_n$). \\

(b) Suppose $\mu_{1,ij}=a_n+b_n$ and $\mu_{2,ij}=a_n-b_n$ with $a_n=o(1)$ and $b_n=o(1)$, $1\leq i<j\leq n$. Then by equation (8), one has
\[\lambda_n=\frac{mn}{2}\frac{b_n^2}{a_n}[1+o(1)].\]
The ratio $\frac{b_n^2}{a_n}$ controls the power, if the sample size $m$ and the number of nodes $n$ are held constant.
Model 1: $a_n=\frac{n^{0.6}}{n}$, $b_n=\sqrt{a_n}$, then $\frac{b_n^2}{a_n}=1$ and $\lambda_n=\frac{mn}{2}[1+o(1)]$.
Model 2: $a_n=\frac{n^{0.7}}{n}$, $b_n=\sqrt{\frac{a_n}{\log n}}$, then $\frac{b_n^2}{a_n}=\frac{1}{\log n}$ and $\lambda_n=\frac{mn}{2
\log n}[1+o(1)]$. Clearly, Model 1 is sparser than Model 2 but our test achieves higher power under Model 1 than Model 2 based on Theorem \ref{thme2}.
\end{Remark}

\begin{Remark}\label{remark}
Recall that the $T_{fro}$ test in \cite{GL18} is defined as
\[T_{fro}=\frac{\sum_{1\leq i<j\leq n}T_{ij}}{t_n},\]
where
\[t_n^2=\sum_{1\leq i<j\leq n}\sum_{k\leq \frac{m}{2}}(A_{G_k,ij}+A_{H_k,ij})\sum_{k> \frac{m}{2}}(A_{G_k,ij}+A_{H_k,ij}).\]
The difference between $\mathcal{T}_n$ and $T_{fro}$ lies in the difference between $s_n$ and $t_n$. Note that $s_n^2$ in $\mathcal{T}_n$ is proved to be a consistent estimator of the variance of $\sum_{1\leq i<j\leq n}T_{ij}$ under $H_0$ for a broad class of distributions $Q$, while $t_n^2$ may not be a consistent estimator of the variance. To see this,
\[\tau_n^2=\mathbb{E}[t_n^2]=\sum_{1\leq i<j\leq n}m^2\mu_{ij}^2.\]
By the proof of Theorem \ref{thmel}, we have $s_n^2=(1+o_p(1))\sigma_n^2$ and
\[
\sigma_n^2=\mathbb{E}[s_n^2]=\sum_{1\leq i<j\leq n}m^2\sigma_{ij}^4.
\]
Here $\sigma_n^2 $ is the variance of $\sum_{1\leq i<j\leq n}T_{ij}$. For any distribution $Q$ with $\tau_n^2\neq (1+o(1))\sigma_n^2$, the test statistic $T_{fro}$ will fail, since $t_n^2$ is not a consistent estimator of $\sigma_n^2$ in this case. For example, let $Q_{ij}(\mu_{ij})$ be the beta distribution $Beta(\alpha,\beta)$. Then for $1\leq i<j\leq n$,
\[\mu_{ij}=\frac{\alpha}{\alpha+\beta},\ \ \sigma_{ij}^2=\frac{\alpha\beta}{(\alpha+\beta)^2(\alpha+\beta+1)}, \ \ \frac{\sigma_{ij}^2}{\mu_{ij}}=\frac{\beta}{(\alpha+\beta)(\alpha+\beta+1)}.\]
When $\alpha$ and $\beta$ are fixed constants, $\tau_n^2\neq (1+o(1))\sigma_n^2$. In this case, $T_{fro}$ doesn't work.

For $Q_{ij}(\mu_{ij})=Bern(\mu_{ij})$, $\sigma_{ij}^2=\mu_{ij}(1-\mu_{ij})$. If $\mu_{ij}\geq 1-\epsilon$ with $\epsilon\in(0,1)$, then $\tau_n^2\neq (1+o(1))\sigma_n^2$. In this case, $T_{fro}$ will fail. On the contrary, when $\mu_{ij}=o(1)$,  $\sigma_{ij}^2=(1+o(1))\mu_{ij}$ and the test $T_{fro}$ may be valid. The simulation results in Section \ref{simu} are consistent with the above findings.
\end{Remark}

\section{Simulation and Real Data}\label{simu}

In this section, we evaluate the finite sample performance of the proposed test $\mathcal{T}_n$ and compare it with the test $T_{fro}$ in \cite{GL18} by simulation. Besides, we apply our test method to a real data.
\subsection{Simulation}
Throughout this simulation, we set the nominal type one error $
\alpha$ to be 0.05. The empirical size and power are obtained by repeating the experiment 1000 times. We take $n=10,30,50,100,200,300$ and $m=2,4,14$.

In the first simulation, we generate weights from beta distribution. Specifically, we generate $G_1,\dots,G_m\sim \mathcal{G}_1=(V,Q,\mu_1)$ with $Q_{ij}(\mu_{1,ij})=Beta(a,b)$ for $1\leq i<j\leq n/2$ or  $n/2<i<j\leq n$ and $Q_{ij}(\mu_{1,ij})=Beta(c,d)$ for $1\leq i\leq n/2<j\leq n$. Denote the graph model as $\mathcal{G}_1(Beta(a,b),Beta(c,d))$.
For a fixed constant $\epsilon$ ($\epsilon\geq 0$), we generate the second sample $H_1,\dots,H_m\sim \mathcal{G}_2=(V,Q,\mu_2)$,  with $Q_{ij}(\mu_{2,ij})=Beta(a+\epsilon,b+\epsilon)$ for $1\leq i<j\leq n/2$ or  $n/2<i<j\leq n$ and $Q_{ij}(\mu_{2,ij})=Beta(c+\epsilon,d+\epsilon)$ for $1\leq i\leq n/2<j\leq n$. Denote the graph model as $\mathcal{G}_2(Beta(a+\epsilon,b+\epsilon),Beta(c+\epsilon,d+\epsilon))$. 

Note that the constant $\epsilon$ ($\epsilon\geq 0$) characterizes the difference between $\mu_{2,ij}$ and $\mu_{1,ij}$ with fixed $a,b,c,d$, since for $Beta(a+\epsilon,b+\epsilon)$, the mean is equal to
\[\mu(\epsilon)=\frac{a+\epsilon-1}{a+b+2\epsilon-2}.\]
Clearly $\mu(\epsilon)$ is an increasing function of $\epsilon$ $(\epsilon\geq 0)$ and larger $\epsilon$ implies larger difference in the means and consequently the power of the test $\mathcal{T}_n$ is supposed to increase. 

We take $a=2,b=3,c=1,d=3$ and $a=9,b=3,c=3,d=2$ to yield right-skewed  and left-skewed beta distributions respectively. The simulation results are summarized in Table \ref{beta1} and Table \ref{beta2}, where the sizes (powers) are reported in column(s) with $\epsilon=0$ ($\epsilon>0$). The sizes and powers of $T_{fro}$ are all zeros, which indicates this test (designed for binary graphs) doesn't apply to weighted graph (see Remark \ref{remark} for explanation). On the contrary, all the sizes of the proposed test $\mathcal{T}_n$ are close to 0.05, which implies the null distribution is valid even for small networks (small $n$) and small sample sizes (small $m$). Besides, the power can approach one, this shows the consistency of the proposed test $\mathcal{T}_n$. The parameter $\epsilon$, $n$, $m$ have significant influence on the powers. As any one of them increases with the rest held constant, the power of $\mathcal{T}_n$ gets higher.

\begin{table}[ht]
	\caption{Simulated size and power with graphs generated from $\mathcal{G}_1(Beta(2,3),Beta(1,3))$ and $\mathcal{G}_2(Beta(2+\epsilon,3+\epsilon),Beta(1+\epsilon,3+\epsilon))$}	
	\centering
	\begin{tabular}{ |p{2cm}|p{1.5cm}|p{2.0cm} p{1.5cm} p{1.5cm}  p{1.5cm}|}
	\hline
	$n(m=2)$      & Method &$\epsilon=0$({\bf size})&  $\epsilon=0.3$\ \ \ ({\bf power}) & $\epsilon=0.5$\ \ \ ({\bf power}) & $\epsilon=0.7$\ \ \ ({\bf power})\\
	\hline
	10  &           &  0.000    &   0.000   &  0.000    & 0.000  \\
	30  &           &  0.000    &  0.000    &  0.000     & 0.000   \\
	50  &           &  0.000    &  0.000   & 0.000     & 0.000  \\
	100   & $T_{fro}$ & 0.000     &  0.000   & 0.000     & 0.000  \\
	200   &          &  0.000    &  0.000   & 0.000     & 0.000  \\
	300   &        &  0.000    & 0.000    & 0.000     & 0.000  \\
	\hline

	10  &           &  0.046    &  0.052   &  0.060    & 0.069   \\
	30  &           &  0.043    &  0.065    & 0.085     & 0.123   \\
	50   &         &  0.049      & 0.079     &  0.093     & 0.203   \\
	100   &        &   0.052    &0.089     &  0.248    & 0.604  \\
	200   & $\mathcal{T}_n$ & 0.045& 0.199   &  0.754     & 0.995  \\
	300   &         &  0.055    & 0.383    &  0.968    & 1.000   \\
	\hline
	\hline
	$n(m=4)$      & Method &$\epsilon=0$({\bf size})&  $\epsilon=0.3$\ \ \ ({\bf power}) & $\epsilon=0.5$\ \ \ ({\bf power}) & $\epsilon=0.7$\ \ \ ({\bf power})\\
	\hline
	10  &           & 0.000     & 0.000  & 0.000      & 0.000   \\
	30  &           & 0.000    & 0.000     & 0.000  &  0.000  \\
	50  &           &  0.000    &  0.000   & 0.000     & 0.000  \\
	100   & $T_{fro}$ & 0.000     &  0.000   & 0.000     & 0.000  \\
	200   &          &  0.000    &  0.000   & 0.000     & 0.000  \\
	300   &        &  0.000    & 0.000    & 0.000     & 0.000  \\
	\hline
	10  &           &  0.049    & 0.058     &  0.065    & 0.069  \\
	30  &           &  0.048    & 0.067    &  0.134     & 0.237   \\
	50   &               &  0.048   &  0.088  & 0.251    & 0.576   \\
	100   & $\mathcal{T}_n$ &  0.056   & 0.209     & 0.757      & 0.992   \\
	200   &               & 0.047       & 0.594     &  0.999     & 1.000   \\
	300   &               & 0.058       & 0.907     &  1.000     & 1.000    \\
	\hline
	\hline
	$n(m=14)$      & Method &$\epsilon=0$({\bf size})&  $\epsilon=0.3$\ \ \ ({\bf power}) & $\epsilon=0.5$\ \ \ ({\bf power}) & $\epsilon=0.7$\ \ \ ({\bf power})\\
	\hline
	10  &           &  0.000    & 0.000     & 0.000     & 0.000  \\
	30  &           &  0.000     & 0.000     & 0.000      & 0.000\\
	50  &           &  0.000     & 0.000     & 0.000     & 0.000   \\
	100   & $T_{fro}$ & 0.000     & 0.000 & 0.000   & 0.006   \\
	200   &          &  0.000     &    0.000  &    0.538   & 1.000  \\
	300   &        &   0.000    & 0.000  & 1.000      &    1.000\\
	\hline
	10  &           &  0.044    &  0.056    &  0.115    & 0.244  \\
	30  &           &  0.048  &  0.193    & 0.707      & 0.980   \\
	50   &        & 0.047     & 0.460 &    0.989  &  1.000   \\
	100   & $\mathcal{T}_n$ &   0.044   &  0.960     &  1.000     &    1.000 \\
	200   &               & 0.041        &  1.000     &    1.000    &  1.000 \\
	300   &               & 0.044        & 1.000      & 1.000   &  1.000    \\
	\hline
\end{tabular}
\label{beta1}
\end{table}

\begin{table}[h]
	\caption{Simulated size and power with graphs generated from $\mathcal{G}_1(Beta(9,3),Beta(3,2))$, $\mathcal{G}_2(Beta(9+\epsilon,3+\epsilon),Beta(3+\epsilon,2+\epsilon))$.}	
	\centering
	\begin{tabular}{ |p{2cm}|p{1.5cm}|p{2.0cm} p{1.5cm} p{1.5cm}  p{1.5cm}|}
	\hline
	$n(m=2)$      & Method &$\epsilon=0$({\bf size})&  $\epsilon=0.5$\ \ \ ({\bf power}) & $\epsilon=0.7$\ \ \ ({\bf power}) & $\epsilon=0.9$\ \ \ ({\bf power})\\
	\hline
	10  &           &  0.000    &   0.000   & 0.000     & 0.000  \\
	30  &           &  0.000    &    0.000  &  0.000     &  0.000 \\
	50  &           &  0.000    & 0.000    & 0.000     & 0.000  \\
	100   & $T_{fro}$ &  0.000     &0.000     & 0.000     & 0.000  \\
	200   &          & 0.000      &0.000     & 0.000     & 0.000  \\
	300   &        &  0.000     &  0.000   &  0.000    & 0.000  \\
	\hline
	10  &           & 0.043     & 0.051     &  0.055    & 0.057  \\
	30  &           &  0.050    &  0.056    &  0.061     & 0.065   \\
	50   &               & 0.047        & 0.062   &  0.076    &  0.077   \\
	100  & $\mathcal{T}_n$ & 0.050      & 0.058   &  0.093    &  0.205   \\
	200   &              &  0.049       & 0.136   & 0.304     & 0.611    \\
	300   &               & 0.057        & 0.231   & 0.591     & 0.926    \\
	\hline
	\hline
	$n(m=4)$      & Method &$\epsilon=0$({\bf size})&  $\epsilon=0.5$\ \ \ ({\bf power}) & $\epsilon=0.7$\ \ \ ({\bf power}) & $\epsilon=0.9$\ \ \ ({\bf power})\\
	\hline
	10  &           & 0.000     & 0.000  & 0.000   & 0.000   \\
	30  &           & 0.000     & 0.000    & 0.000      & 0.000   \\
	50  &           &  0.000    & 0.000    & 0.000     & 0.000  \\
	100   & $T_{fro}$ &  0.000     &0.000     & 0.000     & 0.000  \\
	200   &          & 0.000      &0.000     & 0.000     & 0.000  \\
	300   &        &  0.000     &  0.000   &  0.000    & 0.000  \\
	\hline
	10  &           & 0.047     &  0.053    &  0.056    &  0.057  \\
	30  &           & 0.045     & 0.059     & 0.065      &  0.075  \\
	50   &               & 0.055       & 0.063    & 0.110     & 0.181   \\
	100  & $\mathcal{T}_n$ & 0.053     &0.113     & 0.294     & 0.602   \\
	200   &               & 0.041      & 0.357    & 0.834     &0.989    \\
	300   &                &0.046      & 0.674     & 0.988 & 1.000   \\
	\hline
	\hline
	$n(m=14)$      & Method &$\epsilon=0$({\bf size})&  $\epsilon=0.3$\ \ \ ({\bf power}) & $\epsilon=0.5$\ \ \ ({\bf power}) & $\epsilon=0.7$\ \ \ ({\bf power})\\
	\hline
	10  &           & 0.000     & 0.000  & 0.000    & 0.000   \\
	30  &           &  0.000    & 0.000     & 0.000      & 0.000   \\
	50  &           &   0.000    & 0.000     & 0.000     & 0.000   \\
	100   & $T_{fro}$ & 0.000   & 0.000    &  0.000     & 0.000   \\
	200   &          &  0.000     & 0.000  & 0.000      & 0.000  \\
	300   &        & 0.000      & 0.000     & 0.000      & 0.000   \\
	\hline

	10  &           &  0.044    &   0.060   &   0.064  &  0.077 \\
	30  &           &  0.048   &  0.070    &  0.136     & 0.290   \\
	50   &          & 0.051     &  0.082   & 0.286     &  0.667   \\
	100   & $\mathcal{T}_n$ &  0.049    & 0.187      &  0.784     &  0.998   \\
	200   &               &  0.051       &  0.578     &  1.000      & 1.000  \\
	300   &               & 0.045        &  0.902     &  1.000      &  1.000    \\
	\hline
\end{tabular}
\label{beta2}
\end{table}

In the second simulation, we generate binary graphs to compare the performance of $\mathcal{T}_n$ and $T_{fro}$. Specifically, we generate $G_1,\dots,G_m\sim \mathcal{G}_1=(V,Q,\mu_1)$ with $Q_{ij}(\mu_{1,ij})=Bern(a)$ for $1\leq i<j\leq n/2$ or  $n/2<i<j\leq n$ and $Q_{ij}(\mu_{1,ij})=Bern(b)$ for $1\leq i\leq n/2<j\leq n$. Denote the graph model as $\mathcal{G}_1(Bern(a),Bern(b))$.
For a constant $\epsilon$ ($\epsilon\geq 0$), the second sample $H_1,\dots,H_m$ are generated from $\mathcal{G}_2=(V,Q,\mu_2)$,  with $Q_{ij}(\mu_{2,ij})=Bern(a+\epsilon)$ for $1\leq i<j\leq n/2$ or  $n/2<i<j\leq n$ and $Q_{ij}(\mu_{2,ij})=Bern(b+\epsilon)$ for $1\leq i\leq n/2<j\leq n$. Denote the graph model as $\mathcal{G}_2(Bern(a+\epsilon),Bern(b+\epsilon))$. 
 
We take $a=0.05,b=0.01$, $a=0.1,b=0.05$ and $a=0.5,b=0.5$ to yield sparse, moderately sparse and dense networks respectively. Table \ref{bern1sparse}, Table \ref{bern1} and Table \ref{bern2} summarize the simulation results, where the sizes (powers) are reported in column(s) with $\epsilon=0$ ($\epsilon>0$). When $a=0.05,b=0.01$ and $a=0.1,b=0.05$, the networks are too sparse so that the denominators of $\mathcal{T}_n$ and $T_{fro}$ may be zeros for smaller $n$. Consequently, $\mathcal{T}_n$ and $T_{fro}$ may  not be available and we denote them as NA in 
Table \ref{bern1sparse} and Table \ref{bern1}.

The sizes of $\mathcal{T}_n$ fluctuate around 0.05 and the pattern of powers resemble that in Table \ref{beta1} and Table \ref{beta2}. Since the networks are binary, the test $T_{fro}$ is applicable. For denser networks, the test seems to be pretty conservative since almost all the sizes are less than 0.04 in Table \ref{bern1} and almost all the sizes are zeros in Table \ref{bern2} (see Remark \ref{remark} for explanation). This fact undermines its power 
significantly. On the contrary, the proposed test $\mathcal{T}_n$ has satisfactory power and outperforms $T_{fro}$ substantially. For sparser networks in Table \ref{bern1sparse}, the sizes of $T_{fro}$ are closer to 0.05 and has powers close to that of $\mathcal{T}_n$. This simulation shows the advantage of the proposed test $\mathcal{T}_n$ over $T_{fro}$ under the setting of binary graphs.

\begin{table}[h]
	\caption{Simulated size and power with graphs generated from $\mathcal{G}_1(Bern(0.05),Bern(0.01))$ and $\mathcal{G}_2(Bern(0.05+\epsilon),Bern(0.01+\epsilon))$.}	
	\centering
	\begin{tabular}{ |p{1.9cm}|p{1.3cm}|p{2.0cm} p{1.6cm} p{1.6cm}  p{1.6cm}|}
	\hline
	$n(m=2)$      & Method &$\epsilon=0$({\bf size})&  $\epsilon=0.03$\ \ \ ({\bf power}) & $\epsilon=0.05$\ \ \ ({\bf power}) & $\epsilon=0.07$\ \ \ ({\bf power})\\
	\hline
	10  &           &   NA    &  NA    &   NA   &  NA  \\
	30  &           &   NA    &  NA     &  NA      &  NA    \\
	50  &           &   NA    &  0.038   &  0.094   & 0.236    \\
	100   & $T_{fro}$ &  0.043 & 0.088    &   0.295      &  0.754  \\
	200   &          &   0.044    &  0.244    &  0.880     &  1.000  \\
	300   &        &   0.031     &  0.503     &   0.997     &  1.000   \\
	\hline

	10  &           &  NA     &    NA   &  NA    &    NA \\
	30  &           &  NA     &    NA  & NA       &    NA \\
	50   &           &  NA     &   0.050   &  0.114     & 0.280     \\
	100   & $\mathcal{T}_n$ &  0.052    &  0.099      &  0.343  & 0.791     \\
	200   &               &   0.048 & 0.283 & 0.903       & 1.000   \\
	300   &               &    0.044     &  0.548    & 0.998   & 1.000     \\
	\hline
	\hline
	$n(m=4)$      & Method &$\epsilon=0$({\bf size})&  $\epsilon=0.03$\ \ \ ({\bf power}) & $\epsilon=0.05$\ \ \ ({\bf power}) & $\epsilon=0.07$\ \ \ ({\bf power})\\
	\hline
	10  &           &  NA     &  NA    & NA     & NA   \\
	30  &           &  0.032   & 0.058   & 0.132    & 0.342     \\
	50  &           &   0.043     &  0.078    &  0.314     &  0.750   \\
	100   & $T_{fro}$ &   0.035     & 0.231    &  0.868   & 1.000   \\
	200   &          &  0.049     & 0.740     & 1.000  &  1.000  \\
	300   &        &  0.041      &  0.976     & 1.000    & 1.000    \\
	\hline

	10  &           &   NA    &  NA     & NA      & NA    \\
	30  &           &   0.043    & 0.064     & 0.143       & 0.368    \\
	50   &           &  0.047   &  0.094    &  0.344     & 0.767     \\
	100   & $\mathcal{T}_n$ & 0.050     &  0.259      &   0.885 &  1.000    \\
	200   &               &  0.058    &  0.773      &  1.000   & 1.000   \\
	300   &               &  0.051       &  0.986      & 1.000        &  1.000    \\
	\hline
	\hline
	$n(m=14)$      & Method &$\epsilon=0$({\bf size})&  $\epsilon=0.03$\ \ \ ({\bf power}) & $\epsilon=0.05$\ \ \ ({\bf power}) & $\epsilon=0.07$\ \ \ ({\bf power})\\
	\hline
	10  &           &  NA     &  0.069    &  0.189    & 0.392   \\
	30  &           &  0.038     &   0.263    & 0.872       &  1.000    \\
	50  &           &  0.049      & 0.613     & 1.000      & 1.000    \\
	100   & $T_{fro}$ &  0.040      & 0.995    &  1.000  & 1.000    \\
	200   &          &  0.048     &  1.000    & 1.000      & 1.000   \\
	300   &        &   0.038    &  1.000     & 1.000       &  1.000   \\
	\hline
	10  &           &  NA     & 0.060      & 0.146      & 0.335    \\
	30  &           &   0.048    & 0.277     & 0.858       &  0.998   \\
	50   &           & 0.055      & 0.622     & 1.000      &  1.000    \\
	100   & $\mathcal{T}_n$ &  0.055    & 0.996       & 1.000       &  1.000    \\
	200   &               &  0.060        & 1.000       & 1.000       &  1.000  \\
	300   &               &  0.049       & 1.000       & 1.000        &  1.000    \\
	\hline
\end{tabular}
\label{bern1sparse}
\end{table}

\begin{table}[h]
	\caption{Simulated size and power with graphs generated from $\mathcal{G}_1(Bern(0.1),Bern(0.05))$ and $\mathcal{G}_2(Bern(0.1+\epsilon),Bern(0.05+\epsilon))$.}	
	\centering
	\begin{tabular}{ |p{1.9cm}|p{1.3cm}|p{2.0cm} p{1.6cm} p{1.6cm}  p{1.6cm}|}
	\hline
	$n(m=2)$      & Method &$\epsilon=0$({\bf size})&  $\epsilon=0.03$\ \ \ ({\bf power}) & $\epsilon=0.05$\ \ \ ({\bf power}) & $\epsilon=0.07$\ \ \ ({\bf power})\\
	\hline
	10  &           &  NA    & NA     &    NA  & NA  \\
	30  &           & 0.025     & 0.031     &  0.036     & 0.054   \\
	50   &         & 0.031    & 0.033     & 0.039    &  0.101  \\
	100  &           &0.030     & 0.038    & 0.123    & 0.311 \\
	200   & $T_{fro}$ &0.034     & 0.085    & 0.401     & 0.891        \\
	300   &          &0.046     & 0.157    & 0.756    &  0.998     \\
	\hline
	10  &           & NA     & NA     &    NA  & NA  \\
	30  &           &  0.043    &  0.055    &  0.060     & 0.083   \\
	50   &         & 0.049    & 0.050     & 0.069    & 0.131   \\
	100   &         &0.053     &  0.063   &  0.172   &  0.403 \\
	200   & $\mathcal{T}_n$ &0.049     & 0.123    &  0.483   &  0.933      \\
	300   &          & 0.046   & 0.204    & 0.818    &  0.998     \\
	\hline
	\hline
	$n(m=4)$      & Method &$\epsilon=0$({\bf size})&  $\epsilon=0.03$\ \ \ ({\bf power}) & $\epsilon=0.05$\ \ \ ({\bf power}) & $\epsilon=0.07$\ \ \ ({\bf power})\\
	\hline
	10  &           &   NA  & NA     & NA     & NA  \\
	30  &           & 0.034     &   0.040   &  0.056     & 0.148   \\
	50   &         & 0.030   &  0.037    &  0.120   &   0.328 \\
	100  &           & 0.040     & 0.074    & 0.402     & 0.895  \\
	200   & $T_{fro}$ & 0.031     & 0.277    &     0.962     &  1.000  \\
	300   &          & 0.036      & 0.531    & 1.000     &  1.000 \\
	\hline

	10  &           & NA     & NA     &    NA  & NA  \\
	30  &           & 0.044     & 0.056     & 0.072      & 0.185   \\
	50   &         & 0.049   &  0.053    &  0.165   & 0.392   \\
	100   &        & 0.053      & 0.114    &  0.486    & 0.922  \\
	200   & $\mathcal{T}_n$ &0.049   & 0.339    &     0.976     &  1.000   \\
	300   &         & 0.048      & 0.606     & 1.000     & 1.000   \\
	\hline
	\hline
	$n(m=14)$      & Method &$\epsilon=0$({\bf size})&  $\epsilon=0.03$\ \ \ ({\bf power}) & $\epsilon=0.05$\ \ \ ({\bf power}) & $\epsilon=0.07$\ \ \ ({\bf power})\\
	\hline
	10  &           &  0.039    &  0.044    &  0.075    & 0.156  \\
	30  &           &  0.031    & 0.091     &  0.432     & 0.886    \\
	50  &           &  0.035     & 0.199    &  0.874    & 1.000   \\
	100   & $T_{fro}$ & 0.022      &  0.717   &  1.000     & 1.000   \\
	200   &          &  0.023     &  0.995    &  1.000     &  1.000 \\
	300   &        &  0.026     & 1.000     &  1.000     &  1.000  \\
	\hline
	10  &           &  0.046    & 0.049     & 0.092     & 0.159   \\
	30  &           & 0.055     & 0.120    &  0.480     & 0.889   \\
	50   &           & 0.048     &  0.257   &  0.889    & 1.000    \\
	100   & $\mathcal{T}_n$ &  0.046    &  0.765     &  1.000     & 1.000     \\
	200   &               & 0.040        &  0.997     &  1.000      & 1.000  \\
	300   &               &  0.046       &  1.000     &  1.000      & 1.000    \\
	\hline
\end{tabular}
\label{bern1}
\end{table}

\begin{table}[ht]
	\caption{Simulated size and power with graphs generated from $\mathcal{G}_1(Bern(0.5),Bern(0.4))$ and $\mathcal{G}_2(Bern(0.5+\epsilon),Bern(0.4+\epsilon))$.}	
	\centering
	\begin{tabular}{ |p{1.9cm}|p{1.3cm}|p{2.0cm} p{1.6cm} p{1.6cm}  p{1.6cm}|}
	\hline
	$n(m=2)$      & Method &$\epsilon=0$({\bf size})&  $\epsilon=0.07$\ \ \ ({\bf power}) & $\epsilon=0.10$\ \ \ ({\bf power}) & $\epsilon=0.12$\ \ \ ({\bf power})\\
	\hline
	10  &           &  0.000  & 0.000  & 0.000    & 0.000  \\
	30  &           &  0.000    & 0.000     & 0.000      & 0.002    \\
	50  &          &  0.000    & 0.000     & 0.002    &  0.001 \\
	100   & $T_{fro}$ & 0.000     & 0.000     & 0.005    &  0.022 \\
	200   &          & 0.000     & 0.006     & 0.131    & 0.523  \\
	300   &        & 0.001      & 0.032    & 0.615 & 0.983\\

	\hline
	10  &           &  0.041     & 0.043 &  0.047   & 0.051   \\
	30  &           &  0.048     & 0.054     &   0.077    & 0.101   \\
	50   &         &  0.054    &  0.065    &  0.104   &  0.182 \\
	100   & $\mathcal{T}_n$ & 0.051    & 0.109     & 0.290    & 0.551  \\
	200   &          & 0.050     & 0.298      &   0.797         & 0.981  \\
	300   &          & 0.042     & 0.550    & 0.989  & 1.000\\
	\hline
	\hline
	$n(m=4)$      & Method &$\epsilon=0$({\bf size})&  $\epsilon=0.07$\ \ \ ({\bf power}) & $\epsilon=0.10$\ \ \ ({\bf power}) & $\epsilon=0.12$\ \ \ ({\bf power})\\
	\hline
	10  &           &  0.001    & 0.001     & 0.001     & 0.002  \\
	30  &           &  0.001    & 0.001     &  0.002    & 0.002   \\
	50  &           & 0.001     &0.002     & 0.003     & 0.030  \\
	100   & $T_{fro}$ & 0.000     & 0.007    & 0.139     & 0.502  \\
	200   &          & 0.000     & 0.162    & 0.973     & 1.000   \\
	300   &        &  0.000    &  0.625   &   1.000     & 1.000    \\
	\hline

	10  &           &   0.044   & 0.063     & 0.064     & 0.066  \\
	30  &           &  0.045    &  0.066    &   0.124    &   0.232 \\
	50   &        & 0.058     &0.112     & 0.255     & 0.518  \\
	100   & $\mathcal{T}_n$ & 0.053     & 0.269    & 0.793     & 0.965  \\
	200   &         & 0.047     & 0.787    & 1.000     & 1.000   \\
	300   &          & 0.044     & 0.983    &  1.000      &  1.000   \\
	\hline
	\hline
	$n(m=14)$      & Method &$\epsilon=0$({\bf size})&  $\epsilon=0.07$\ \ \ ({\bf power}) & $\epsilon=0.10$\ \ \ ({\bf power}) & $\epsilon=0.12$\ \ \ ({\bf power})\\
	\hline
	10  &           &  0.000   & 0.001     & 0.002    & 0.014  \\
	30  &           &  0.000   & 0.013     & 0.174     & 0.559   \\
	50  &           &  0.000    & 0.083    & 0.812     & 0.997   \\
	100   & $T_{fro}$ & 0.000      & 0.834    & 1.000      & 1.000   \\
	200   &          &  0.032     & 0.998     &  1.000     & 1.000  \\
	300   &        &   0.032    &  1.000    & 1.000      & 1.000   \\
	\hline
	10  &           & 0.052    & 0.075     & 0.120     & 0.189  \\
	30  &           & 0.040    & 0.273     & 0.741      & 0.955   \\
	50   &           & 0.040  & 0.648  & 0.990    &  1.000  \\
	100   & $\mathcal{T}_n$ &  0.055    & 0.997      & 1.000      &    1.000 \\
	200   &               &  0.049       & 0.998      & 1.000 & 1.000   \\
	300   &               &  0.050       &  1.000     &  1.000      &  1.000    \\
	\hline
\end{tabular}
\label{bern2}
\end{table}

\subsection{Real Data Application}
In this section, we consider applying the proposed method to a real life data that can be downloaded from a public database \url{(http://fcon_1000.projects.nitrc.org/indi/retro/cobre.html)}. This data set contains Raw anatomical and functional
scans from 146 subjects (72 patients with schizophrenia and 74 healthy controls). After a series of processes done by \cite{AR19}, only 124 subjects (70 patients with schizophrenia and 54 healthy controls) were kept. In their study, 263 brain regions of interests were chosen as nodes and connectivity between nodes were measured by the edge weights that represent the Fisher-transformed correlation between the fMRI time series of the nodes after passing to ranks \cite{JD19}.

As the healthy group (54 networks) and patient group (70 networks) have different sample sizes, our test statistic is not directly applicable. We adopt the following two methods to solve this issue. The first way is to randomly sample 16 networks from the 54 networks in the health group and unite them with the 54 networks of health group to yield 70 samples. Then the healthy group and patient group have equal sample sizes and we can calculate the test statistics. This process is repeated 100 times and the five-number summary of the test statistics is presented in Table \ref{real2}. The second method is to randomly sample 54 networks from the schizophrenia patient group and then calculate the test statistics based on the sampled 54 networks and the 54 networks in the healthy group. The random sampling procedure is repeated 100 times and the five-number summary of test statistics is presented in Table \ref{real1}. All the calculated test statistics $\mathcal{T}_n$ and $T_{fro}$ are much larger than 1.96, which leads to the same conclusion that the patient population significantly differs from the healthy population at significance level $\alpha=0.05$. Moreover, the proposed test statistic $\mathcal{T}_n$ is almost twice of $T_{fro}$, implying that our test is more powerful to detect the population difference.

The computation of the proposed test statistic requires randomly splitting two samples into two groups. 
In order to evaluate the effect of the random splitting on the proposed test, we randomly sample 54 networks from the patient group, denoted as $G_k,(1\leq k\leq 54)$. Let $H_k, (1\leq k\leq 54)$ be the 54 networks in the healthy group. Consider $G_k,H_k, (1\leq k\leq 54)$ as the two samples. We randomly partition the two samples into two groups, denoted as $\tilde{G}_k,\tilde{H}_k, (1\leq k\leq 27)$ and  $\tilde{G}_k,\tilde{H}_k, (27<k\leq 54)$ and then compute the test statistics $\mathcal{T}_n$ and $T_{fro}$. This procedure is repeated 100 times and the five-number summary of the 100 calculated test statistics are recorded in Table \ref{real3}. The same conclusion could be drawn based on the 100 statistics, implying that random splitting doesn't significantly affect the proposed method.

Additionally, to compare the performance of $\mathcal{T}_n$ and $T_{fro}$ in binary graph setting, we artificially transform the weighted graphs to binary graphs by thresholding as follows. For a given threshold $\tau$, if the absolute value of an edge weight is greater (smaller) than $\tau$, then the edge is transformed to 1 (0). Smaller (larger) $\tau$ yields denser (sparser) networks. We take the threshold values $\tau\in\{0.01, 0.03, 0.1, 0.3, 0.5, 0.7, 0.9\}$. For each $\tau$, we calculate the test statistics as in Table \ref{real2} and the results are summarized in Table \ref{real4}. The threshold $\tau$ dramatically affects the conclusion. For $0.1\leq\tau\leq 0.7$, both $\mathcal{T}_n$ and $T_{fro}$ reject the null hypothesis $H_0$ that the two network populations are the same, with $\mathcal{T}_n$ more powerful than $T_{fro}$ in most cases. However, for $\tau=0.01$ (denser networks), $\mathcal{T}_n$ rejects the null hypothesis $H_0$, while $T_{fro}$ fails to reject $H_0$. 
This analysis outlines the importance to develop testing procedures for weighted networks, as artificial transforming weighted networks to unweighted networks may lead to contradictory conclusions.

\begin{table}[h]
	\caption{Repeat sampling 16 networks from 54 networks in the healthy group.}	
	\centering
	\begin{tabular}{ |p{2cm} | p{1.5cm} p{2cm} p{2cm} p{2cm}  p{2cm}|}
	\hline
 Method & Min. &  1st Qu. & Median & 3rd Qu. & Max.\\
	\hline
  $\mathcal{T}_n$   &   43.52 &   50.33 &   53.62 &   57.05 &   62.39    \\
  	\hline
   $T_{fro}$        &  22.27&   25.93&   27.43  &  29.45 &  32.34      \\
	\hline
\end{tabular}
\label{real2}
\end{table}

\begin{table}[h]
	\caption{Repeat sampling 54 networks from 70 networks in patient group.}	
	\centering
	\begin{tabular}{ |p{2cm} | p{1.5cm} p{2cm} p{2cm} p{2cm}  p{2cm}|}
	\hline
 Method & Min. &  1st Qu. & Median & 3rd Qu. & Max.\\
	\hline
  $\mathcal{T}_n$   &   27.81 &   35.45 &   37.55   &   39.83   & 47.57    \\
  	\hline
   $T_{fro}$        &  12.33 &   15.09 &   15.88   &   16.95   & 20.53 \\
	\hline
\end{tabular}
\label{real1}
\end{table}

\begin{table}[h]
	\caption{Random splitting of two samples $G_k,H_k, 1\leq k\leq 54$.}	
	\centering
	\begin{tabular}{ |p{2cm} | p{1.5cm} p{2cm} p{2cm} p{2cm}  p{2cm}|}
	\hline
 Method & Min. &  1st Qu. & Median & 3rd Qu. & Max.\\
	\hline
  $\mathcal{T}_n$   &  28.25 &  35.28 &  38.19   &   40.95 &  46.33    \\
  	\hline
   $T_{fro}$        &  12.30 &  15.12 &  16.12  &  17.18  & 19.21\\
	\hline
\end{tabular}
\label{real3}
\end{table}

\begin{table}[h]
	\caption{Transforming weighted graphs to unweighted graphs with different threshold $\tau$.}	
	\centering
	\begin{tabular}{ |p{4cm} | p{1.5cm} p{1.5cm} p{1.5cm} p{1.5cm}  p{1.5cm}|}
	\hline
 Method & Min. &  1st Qu. & Median & 3rd Qu. & Max.\\
	\hline
  $\mathcal{T}_n$\ \ \ ($\tau=0.01$)  & 9.70 &16.40 & 19.61 &22.10& 31.88 \\
  	\hline
   $T_{fro}$ ($\tau=0.01$)       & 0.42 & 0.70 &   0.83& 0.93 & 1.31\\
	\hline
	\hline
  $\mathcal{T}_n$ \ \ ($\tau=0.03$)  &  11.91 & 17.32 & 21.20 & 24.90& 33.23 \\
  	\hline
   $T_{fro}$ ($\tau=0.03$)       &   1.54 & 2.17 &   2.62 & 3.05 & 3.95\\
	\hline
	\hline
  $\mathcal{T}_n$ \ \  ($\tau=0.1$)  &  11.86&    19.66&    23.01 &     25.78 &   38.03 \\
  	\hline
   $T_{fro}$  ($\tau=0.1$)       &  4.88 &    7.81 &    9.01 &       9.93 &  14.10\\
	\hline
	\hline
  $\mathcal{T}_n$ \ \  ($\tau=0.3$)  &  14.06 &  26.37&   29.64   &  32.42 &  41.33 \\
  	\hline
   $T_{fro}$ ($\tau=0.3$)       &  11.81 &  20.90 &  23.20   &   25.34&   32.16 \\
    	\hline
	\hline
  $\mathcal{T}_n$ \ \  ($\tau=0.5$)  &  11.86 &  16.57 &  18.19   &  19.21 &  24.43  \\
  	\hline
   $T_{fro}$ ($\tau=0.5$)       &  10.14 &  13.81&   15.10   &   16.14 &  20.14 \\
    	\hline
	\hline
  $\mathcal{T}_n$ \ \  ($\tau=0.7$)  &  5.60 &   6.93 &   7.32   &   7.93 &   9.37 \\
  	\hline
   $T_{fro}$ ($\tau=0.7$)       & 6.55  &  8.02 &   8.53   &   9.08 &  10.60  \\
    	\hline
	\hline
  $\mathcal{T}_n$ \ \ ($\tau=0.9$)  & 0.47 & 1.79 & 2.19  &  2.52 & 3.36\\
  	\hline
   $T_{fro}$  ($\tau=0.9$)      &  0.81 & 2.54 & 2.89  &  3.55&  5.02 \\
    	\hline
\end{tabular}
\label{real4}
\end{table}


\section{Proof of Main Results}\label{proof}

\medskip

\noindent
\textit{Proof of Theorem \ref{thmel}:} We employ the Lindeberg Central Limit Theorem to prove theorem \ref{thmel}.

Firstly, note that under $H_0$, we have
\begin{eqnarray*}
\sigma_n^2&=&\mathbb{E}[s_n^2]=\sum_{1\leq i<j\leq n}\mathbb{E}[T_{ij}^2]\\
&=&\sum_{1\leq i<j\leq n}\mathbb{E}\Big[\sum_{k\leq \frac{m}{2}}(A_{G_k,ij}-A_{H_k,ij})\Big]^2\mathbb{E}\Big[\sum_{k> \frac{m}{2}}(A_{G_k,ij}-A_{H_k,ij})\Big]^2\\
&=&\sum_{1\leq i<j\leq n}\sum_{k\leq \frac{m}{2}}\mathbb{E}(A_{G_k,ij}-A_{H_k,ij})^2\sum_{k> \frac{m}{2}}\mathbb{E}(A_{G_k,ij}-A_{H_k,ij})^2\\
&=&\sum_{1\leq i<j\leq n}m^2\sigma_{ij}^4,
\end{eqnarray*}

Next, we verify the Lindeberg condition. By Cauchy-Schwarz inequality and Markov inequality, it follows that for any $\epsilon>0$,
\begin{eqnarray*}
\mathbb{E}T_{ij}^2I[|T_{ij}|>\epsilon \sigma_n]&\leq &\sqrt{\mathbb{E}T_{ij}^4\mathbb{P}[|T_{ij}|>\epsilon \sigma_n]}
\leq  \sqrt{\mathbb{E}T_{ij}^4\frac{\mathbb{E}T_{ij}^4}{\epsilon^4 \sigma_n^4}}=\frac{\mathbb{E}T_{ij}^4}{\epsilon^2 \sigma_n^2}.
\end{eqnarray*}
Notice that
\begin{eqnarray*}
\mathbb{E}T_{ij}^4&=&\sum_{k_1,k_2,k_3,k_4\leq \frac{m}{2}}\mathbb{E}(A_{G_{k_1},ij}-A_{H_{k_1},ij})(A_{G_{k_2},ij}-A_{H_{k_2},ij})(A_{G_{k_3},ij}-A_{H_{k_3},ij})(A_{G_{k_4},ij}-A_{H_{k_4},ij})\\
&\times & \sum_{k_1,k_2,k_3,k_4> \frac{m}{2}}\mathbb{E}(A_{G_{k_1},ij}-A_{H_{k_1},ij})(A_{G_{k_2},ij}-A_{H_{k_2},ij})(A_{G_{k_3},ij}-A_{H_{k_3},ij})(A_{G_{k_4},ij}-A_{H_{k_4},ij}).
\end{eqnarray*}
Since for distinct $k_1$, $k_2$, $k_3,k_4\in\{1,2,\dots,m\}$, \[\mathbb{E}[(A_{G_{k_1},ij}-A_{H_{k_1},ij})^3(A_{G_{k_2},ij}-A_{H_{k_2},ij})]=0,\]
\[\mathbb{E}[(A_{G_{k_1},ij}-A_{H_{k_1},ij})^2(A_{G_{k_2},ij}-A_{H_{k_2},ij})(A_{G_{k_3},ij}-A_{H_{k_3},ij})]=0,\]
\[\mathbb{E}[(A_{G_{k_1},ij}-A_{H_{k_1},ij})(A_{G_{k_2},ij}-A_{H_{k_2},ij})(A_{G_{k_3},ij}-A_{H_{k_3},ij})(A_{G_{k_4},ij}-A_{H_{k_4},ij})]=0.\]
As a result, it follows
\begin{eqnarray*}
&&\sum_{k_1,k_2,k_3,k_4\leq \frac{m}{2}}\mathbb{E}(A_{G_{k_1},ij}-A_{H_{k_1},ij})(A_{G_{k_2},ij}-A_{H_{k_2},ij})(A_{G_{k_3},ij}-A_{H_{k_3},ij})(A_{G_{k_4},ij}-A_{H_{k_4},ij})\\
&=&\sum_{k_1=k_2\neq k_3=k_4\leq \frac{m}{2}}\mathbb{E}(A_{G_{k_1},ij}-A_{H_{k_1},ij})^2(A_{G_{k_3},ij}-A_{H_{k_3},ij})^2\\
&&+\sum_{k_1=k_3\neq k_2=k_4\leq \frac{m}{2}}\mathbb{E}(A_{G_{k_1},ij}-A_{H_{k_1},ij})^2(A_{G_{k_2},ij}-A_{H_{k_2},ij})^2\\
&&+\sum_{k_1=k_4\neq k_2=k_3\leq \frac{m}{2}}\mathbb{E}(A_{G_{k_1},ij}-A_{H_{k_1},ij})^2(A_{G_{k_2},ij}-A_{H_{k_2},ij})^2\\
&&+\sum_{k_1=k_4= k_2=k_3\leq \frac{m}{2}}\mathbb{E}(A_{G_{k_1},ij}-A_{H_{k_1},ij})^4\\
&=&\sum_{k_1=k_4= k_2=k_3\leq \frac{m}{2}}\mathbb{E}(A_{G_{k_1},ij}-A_{H_{k_1},ij})^4+3\sum_{k_1\neq k_2\leq \frac{m}{2}}\mathbb{E}(A_{G_{k_1},ij}-A_{H_{k_1},ij})^2(A_{G_{k_2},ij}-A_{H_{k_2},ij})^2
\end{eqnarray*}

Then we have
\begin{eqnarray*}
\mathbb{E}T_{ij}^4&=&\Big[\sum_{k\leq \frac{m}{2}}\mathbb{E}(A_{G_{k},ij}-A_{H_{k},ij})^4+3\sum_{k_1\neq k_2\leq \frac{m}{2}}\mathbb{E}(A_{G_{k_1},ij}-A_{H_{k_1},ij})^2(A_{G_{k_2},ij}-A_{H_{k_2},ij})^2\Big]\\
&\times & \Big[\sum_{k> \frac{m}{2}}\mathbb{E}(A_{G_{k},ij}-A_{H_{k},ij})^4+3\sum_{k_1\neq k_2> \frac{m}{2}}\mathbb{E}(A_{G_{k_1},ij}-A_{H_{k_1},ij})^2(A_{G_{k_2},ij}-A_{H_{k_2},ij})^2\Big]\\
&=&\Big(m^2\sigma_{ij}^4+3m\eta_{ij}\Big)^2=m^4\sigma_{ij}^8+6m^3\sigma_{ij}^4\eta_{ij}+9m^2\eta_{ij}^2,
\end{eqnarray*}
where $\eta_{ij}=\mathbb{E}(A_{G_{k},ij}-A_{H_{k},ij})^4$. Hence, by condition (\ref{cond}), it follows that
\begin{eqnarray*}
&&\frac{1}{\sigma_n^2}\sum_{1\leq i<j\leq n}\mathbb{E}T_{ij}^2I[|T_{ij}|>\epsilon \sigma_n]\leq \frac{1}{\epsilon^2\sigma_n^4}\sum_{1\leq i<j\leq n}\mathbb{E}T_{ij}^4\\
&\leq &\frac{m^4\sum_{1\leq i<j\leq n}\sigma_{ij}^8+6m^3\sum_{1\leq i<j\leq n}\sigma_{ij}^4\eta_{ij}+9m^2\sum_{1\leq i<j\leq n}\eta_{ij}^2}{\epsilon^2m^4(\sum_{1\leq i<j\leq n}\sigma_{ij}^4)^2}\rightarrow0.
\end{eqnarray*}
By the Lindeberg Central Limit Theorem, we conclude that $\sigma_n^{-1}\sum_{1\leq i<j\leq n}T_{ij}$ converges in distribution to $N(0,1)$.

Finally, we prove $\mathcal{T}_n$  converges in distribution to $N(0,1)$ by proving that $s_n^2=(1+o_p(1))\sigma_n^2$. Note that for $i<j$ and $k<l$,
\[
\mathbb{E}\big[(T_{ij}^2-m^2\sigma_{ij}^4)(T_{kl}^2-m^2\sigma_{kl}^4)\big]=0, \hskip 1cm if\ \{i,j\}\neq\{k,l\}.
\]
 Consequently, one has
\begin{eqnarray*}
\mathbb{E}\big[s_n^2-\sigma_n^2\big]^2&=&\mathbb{E}\big[ \sum_{1\leq i<j\leq n}(T_{ij}^2-m^2\sigma_{ij}^4)\big]^2= \sum_{1\leq i<j\leq n}\mathbb{E}\big[(T_{ij}^2-m^2\sigma_{ij}^4)\big]^2=O(n^2m^4),
\end{eqnarray*}
Hence, $s_n^2=\sigma_n^2+O_p(\sqrt{n^2}m^2)$. If $\sqrt{n^2}m^2=o(\sigma_n^2)$, then $s_n^2=(1+o_p(1))\sigma_n^2$, which implies $\mathcal{T}_n$ converges in distribution to $N(0,1)$ by Slutsky's theorem.

\qed

\noindent
\textit{Proof of Theorem \ref{thme2}:}

Under $H_1$, we have
\begin{eqnarray*}
\Lambda_{ij}=\mathbb{E}[T_{ij}]
&=&\mathbb{E}\Big[\sum_{k\leq \frac{m}{2}}(A_{G_k,ij}-A_{H_k,ij})\Big]\mathbb{E}\Big[\sum_{k> \frac{m}{2}}(A_{G_k,ij}-A_{H_k,ij})\Big]\\
&=&\sum_{k\leq \frac{m}{2}}\mathbb{E}(A_{G_k,ij}-A_{H_k,ij})\sum_{k> \frac{m}{2}}\mathbb{E}(A_{G_k,ij}-A_{H_k,ij})\\
&=&\frac{m^2}{4}(\mu_{1,ij}-\mu_{2,ij})^2,
\end{eqnarray*}
and
\begin{eqnarray*}
V_{ij}=\mathbb{E}(A_{G_k,ij}-A_{H_k,ij})^2
&=&\mathbb{E}(A_{G_k,ij}-\mu_{1,ij}+\mu_{1,ij}-\mu_{2,ij}+\mu_{2,ij}-A_{H_k,ij})^2\\
&=&\mathbb{E}(A_{G_k,ij}-\mu_{1,ij})^2+\mathbb{E}(\mu_{1,ij}-\mu_{2,ij})^2+\mathbb{E}(\mu_{2,ij}-A_{H_k,ij})^2\\
&=&\sigma_{1,ij}^2+\sigma_{2,ij}^2+(\mu_{1,ij}-\mu_{2,ij})^2.
\end{eqnarray*}
Then
\[\sigma_{1,n}^2=\mathbb{E}s_n^2=\frac{m^2}{4}\sum_{1\leq i<j\leq n}V_{ij}^2,\]
and
\begin{eqnarray*}
\mathbb{E}\big[s_n^2-\sigma_{1,n}^2\big]^2&=&\mathbb{E}\big[ \sum_{1\leq i<j\leq n}(T_{ij}^2-\frac{m^2}{4}V_{ij}^2)\big]^2= \sum_{1\leq i<j\leq n}\mathbb{E}\big[(T_{ij}^2-\frac{m^2}{4}V_{ij}^2)\big]^2=O(n^2m^4).
\end{eqnarray*}
Hence, $s_n^2=\sigma_{1,n}^2(1+o_P(1))$ if $nm=o(m^4\sum_{1\leq i<j\leq n}V_{ij}^2)$. 
Note that 
\[\mathbb{E}\big[\sum_{1\leq i<j\leq n}(T_{ij}-\Lambda_{ij})\big]^2=\sum_{1\leq i<j\leq n}\mathbb{E}(T_{ij}-\Lambda_{ij})^2\leq \sigma_{1,n}^2.\]
As a result, under $H_1$, the test statistic is decomposed as
\begin{eqnarray*}
\mathcal{T}_n&=&\frac{ \sum_{1\leq i<j\leq n}(T_{ij}-\Lambda_{ij})}{s_n}+\frac{ \sum_{1\leq i<j\leq n}\Lambda_{ij}}{s_n}\\
&=&\frac{ \sigma_{1,n}}{s_n}\Big(\frac{ \sum_{1\leq i<j\leq n}(T_{ij}-\Lambda_{ij})}{\sigma_{1,n}}+\frac{ \sum_{1\leq i<j\leq n}\Lambda_{ij}}{\sigma_{1,n}}\Big)\\
&=&\frac{ \sum_{1\leq i<j\leq n}\Lambda_{ij}}{\sigma_{1,n}}+O_P(1).
\end{eqnarray*}

\section*{Acknowledgement}
The authors are grateful to the Editor, the Associate Editor and Referees for helpful comments that
significantly improved this manuscript.

\end{document}